# Analysis and Clustering of Workload in Google Cluster Trace based on Resource Usage

Mansaf Alam
Department of Computer Science
Jamia Millia Islamia
New Delhi, India
malam2@jmi.ac.in

Kashish Ara Shakil
Department of Computer Science
Jamia Millia Islamia
New Delhi, India
shakilkashish@yahoo.co.in

Shuchi Sethi
Department of Computer Science
Jamia Millia Islamia
New Delhi, India
shuchi.sethi@yahoo.com

*Abstract*—Cloud computing has gained interest amongst commercial organizations, research communities, developers and other individuals during the past few years. In order to move ahead with research in field of data management and processing of such data, we need benchmark datasets and freely available data which are publicly accessible. Google in May 2011 released a trace of a cluster of 11k machines referred as "Google Cluster Trace". This trace contains cell information of about 29 days. This paper provides analysis of resource usage and requirements in this trace and is an attempt to give an insight into such kind of production trace similar to the ones in cloud environment. The major contributions of this paper include Statistical Profile of Jobs based on resource usage, clustering of Workload Patterns and Classification of jobs into different types based on k-means clustering. Though there have been earlier works for analysis of this trace, but our analysis provides several new findings such as jobs in a production trace are trimodal and there occurs symmetry in the tasks within a long job type.

*Keywords—cloud computing; Google cluster trace; k-means clustering*

## I. INTRODUCTION

There has been a tremendous increase of interest amongst commercial organizations, research communities, developers and other individuals and organization in cloud computing during the past few years. Thousands of petabytes of data is being generated by different sources such as sensors, social networking websites, and data streaming and tracking of transaction histories. Cloud computing is now slowly emerging as a desirable and preferred platform for management of such kind of data. Cloud computing provides several benefits such as dynamic scalability, rapid elasticity and pay per use feature. Though cloud computing seems to be a very lucrative and attractive approach for management of such kind of data and several companies such as Google, Amazon and Yahoo are using techniques such as Big table and MapReduce based Hadoop Framework for management of data of such huge volume, variety, velocity and veracity but still research on data management of such data is still not mature. Cloud Database Management System Architecture [12] is architecture for management of data in cloud environment.

In order to move ahead with research in field of data management and processing of such data, we need benchmark datasets and freely available data which are publicly accessible. Even though there is a growing need for universal accessibility of such datasets but still such data is not easily available. There might be several reasons for it such as confidentiality of organizations data, confidentiality of client's data and organizational policies. Furthermore, addressing and resolving challenges of such distributed and huge datasets through cloud computing require a detailed understanding of workload behaviour of cloud computing based on changes in resource requirements.

Google in May 2011 released a trace of a cluster of 11k machines referred as "Google Cluster Trace". This trace contains cell information of about 29 days [1]. The cloud trace data is highly anonymous [2] in terms of task, job names and timestamps being anonymous. This paper provides an analysis and clustering of workload in Google cluster trace which varies with variations in resource requirements. The major contributions of this paper are:

- Statistical Profile of Jobs based on resource usage
- Clustering of Workload Patterns
- Classification of jobs into different types based on k-means clustering.

The remainder of this section is organized as follows: In Section II we provide an insight about the prior work which has been done for characterization and analysis of Google cluster data. Section III provides an insight into Google cluster trace. Moving further Section IV provides statistical profile of job. In Section V clustering of workload patterns have been performed. Finally we conclude this paper with conclusion and future directions in section VI.

## II. RELATED WORK

There have been prior studies related to characterization of workloads in single servers but during the past few years study about workload analysis of multiple servers' has also gained momentum. Many attempts are now being made to analyse workloads of such kind of setups particularly the ones in cloud. Study of Google cluster trace is the stepping stone towards analysis of workload in a setup with multiple servers or cloud environment.

Chen et al. [2] have analysed Google cluster trace and have provided a statistical profile of the data sets based on their temporal behaviour which reveals that each job has different behaviour with contrasting arrival rates but continuous allocations. Clustering analysis has also been performed by them for identifying group of common jobs using k means clustering approach.

In [3] the authors have studied Google cluster trace data and defined how the machines are managed in the trace. They have shown that 95% machines have 0.5 as the memory capacity values. Their results also reveal that each machine has a status update of more than 30 times and there exists a temporal locality of machine events. They have also explored the workload behaviour of the jobs, according to them 40.52% of the jobs that are scheduled are killed at least once in a lifetime. As per their findings frequency with which a scheduled job is killed is higher than its frequency when it fails. The Scheduling constraints results reveal that 93% jobs have less than four constraints and 98% jobs have less than 20 constraints. Furthermore they have also analysed temporal level usage of resources such as CPU usage.

Reiss et al. [4] have described the heterogeneity of the resources in Google cluster. They have notified that there is a wide heterogeneity in the different resources which are available and includes RAM, CPU core, memory usage, cache memory and page memory usage and there also exists heterogeneity in the manner in which resources are utilized. They have analyzed that due to the resources being highly heterogeneous in nature, the resource usage pattern is predictable but this prediction is weak. The paper also shows that most of the jobs in the workload are repeated and resource requirements of longer jobs are stable. In [5] a DMMM framework has been proposed which also works on the prediction of the resource usage in cloud environment. This prediction is refined further in [13] where a network latency based approach is used.

In [6] host workload has been predicted using Google cluster, their prediction method is based on Bayes model. They have experimentally used Bayes method to show that 56.50 % accuracy is achieved for predicting host load. BiDAi [7] is a log data analysis framework that uses big data technologies for analysis of large cluster traces. It imports data in CSV format and uses SQL Lite and Hadoop distributed file system at the backend. It supports a mixture of R language and Hadoop MapReduce commands. According to [8] jobs/event follow a joint ratio of approx. 10% and < 2% for resource utilization per application. And they have also concluded a Pareto similar distribution on the number of application per set using k means clustering approach. Mishra et al. [9] have deduced that tasks in the cluster are either of short duration or of long duration i.e. they are bimodal in nature but most of the tasks are short duration ones. According to them resources are consumed most of the times by longer duration's tasks.

This manuscript incorporates certain observations which showed up during the experiments that are already part of certain related work in [2][3][4].The work in this paper takes a novel aspect of resources and studies job behaviour with respect to resource availability and usage.

## III. GOOGLE CLUSTER TRACE DATA

Google cluster trace [1] is a dataset released by Google in May 2011 and contains cell information of about 29 days. In this paper ClusterData2011_1 has been used. Each cell represents a set of several machines sharing a single cluster management system. Each job in trace contains one or many tasks where each task might contain several processes that are to be run on a single machine. The trace data contains about five tables: Machine Events Table, Machine attribute's Table, Jobs Events Table, Task Events Table, Task Constraints Table and Task Resource Usage Table. Following is a brief description of each of the tables:

### A. Machine Events Table

Machine Event table contains a description of each of the machines. It contains information about the timestamp at which machine was started, the ID of the machine, the various events type which can be ADD (0),REMOVE (1) and UPDATE(2) Platform ID, CPU and memory capacity. Thus machine capacity has two dimensions CPU capacity and size of RAM.

On analysis of this table it was revealed that about 92.651% ~ approx. 93% have a CPU capacity of 0.5. Subsequently after analysis of the memory capacity four different broad levels were found: '0.25', '0.50', '0.75' and '1'. Table I shows the percentage of machines at each memory capacity levels.

TABLE I. PERCENTAGE OF MACHINES AT EACH MEMORY CAPACITY LEVEL

| Memory Capacity | Percentage of Machines |
|---|---|
| 0.25 | 27% |
| 0.50 | 57.5% |
| 0.75 | 8% |
| 1 | 6% |

### B. Machine Attributes Table

Machine attributes are represented as a key value pair and describe properties of machine for example clock speed and version of machine kernel. It is represented by five columns which include timestamp, ID of machine, name and value of attribute. It also contains information represented by a Boolean value about whether an attribute was deleted or not.

### C. Jobs Events Table

There are several events encountered by job or a task during its lifetime. They may include events such as submission of a job, its scheduling, eviction on being rescheduled, failure, completion or an event where a job might be killed and many others. This table contains fields like timestamp, a field in missing information, type of event, a scheduling class which represents latency sensitivity of a job or task, name of job which is mostly unique and generated in a hashed manner by programs such as MapReduce.

### D. Task Events Table

Task events table contains information about various events pertaining to different tasks. This table is represented by thirteen columns which include fields such as time stamp, missing information, index of tasks, event type, priority of a task which can be free priorities or a production priority or perhaps a monitoring priority, information about resource requests and machine constraint's.

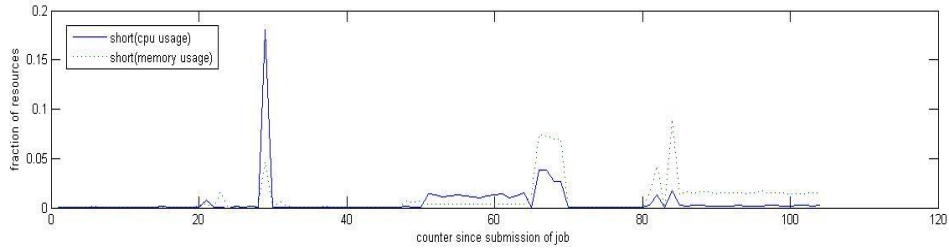

Fig.1.Resource Usage in a Short Job

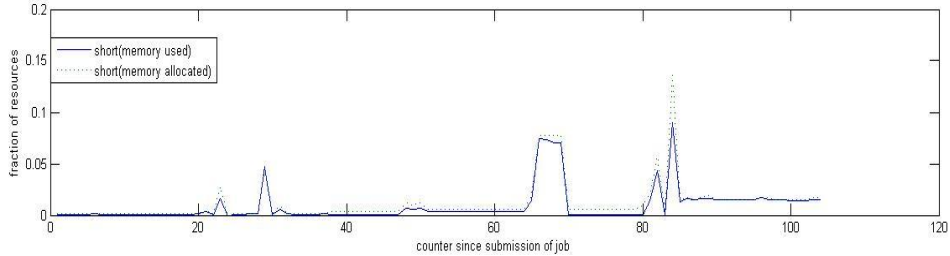

Fig.2.Resource allocated and used in a Short Job

*E. Task Constraints Table*

Each task may be governed by one or more constraints. Each machine attribute can have several constraints. The constraint table contains task indices, their attributes followed by the attribute name, attribute value and a comparison operator. These comparison operators can be LESS THAN, GREATER THAN, EQUAL AND NOT EQUAL

*F. Task Resource Usage Table*

This table contains a report of resource usage for every 5 minutes. These measurements are taken at an interval of every 1 second. It contains start and end time, task index, CPU rate, memory usage information, amount of assigned memory, cache memory usage, disk I/O time, sampling rate, cycles, MAI. Local disk space usage and aggregation type as the fields.

IV. STATISTICAL PROFILE OF JOBS

Analysis of the cluster trace revealed several interesting outcomes. Our analysis is based on observation of one day's data from the trace dataset. These observations can also be generalised for other production cluster traces similar to a cloud environment. It shows the behaviour of jobs with respect to time and resource usage.

After reviewing the literature it was revealed that duration for which tasks are executed is bimodal [3] [9] carrying this notion further, upon analysis of the trace we could deduce that jobs are trimodal in nature i.e. Jobs can be categorized into three major types based on their resource requirements and the number of times an event request is made. These categories are :( 1) Long jobs, (2) Short jobs and (3) medium jobs. We could further sub categorize resource usage pattern by each of the job types as (1) less resource usage, (2) mid resource usage and (3) resource hungry. Less resource usage is the jobs that don't require or use many resources such as CPU and memory usage. Mid resource usage represents those jobs that require a fair amount of resources but are not resource hungry. The third category is resource hungry jobs. These are the jobs that require a lot of resources and are always on the hunt for more resources. Further analysis of each of the job types is revealed in the following sections:

*A. Short Jobs*

These are the jobs that are very short in duration as compared to other jobs in the trace. Fig.1. represents information about resource usage in such a job type. For the purpose of analysis we have used two major resources memory and CPU only. This figure shows that the resource requirement in a short term job is less and if at all it increases, it increases in bursts which can be indicated by sudden spikes in the figure. Fig. 2. Represents a graph between memory allocated and memory used. From analysis of this graph it can be observed that all the resources that were allocated were well used, it can be concluded due to a match between the resource allocated and their usage behaviour. It addition to this it can also be inferred that these short term jobs are never resource hungry and have low resource requirement as well as usage.

*B. Medium Jobs*

These are the jobs that are slightly medium i.e. they are neither as small as short jobs nor as big as long jobs. They do occur frequently but lesser number of times than short jobs but more than long jobs. Fig. 3. Symbolizes resource usage behaviour of a medium job. This figure reveals that most of the requirements for resources were for memory and hence such jobs can be classified as memory intense. This figure also shows that resources were used in the best possible manner by the medium jobs. Fig.4.represents a graph between memory allocated and memory used. It can be observed from this figure that most of the resource requirements were not only well addressed.

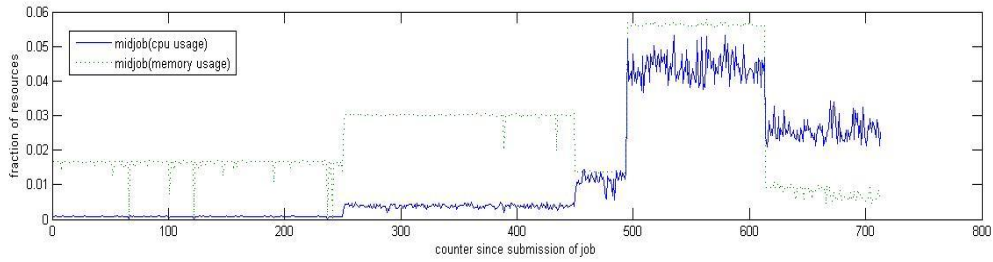

Fig. 3. Resource Usage in a Medium Job

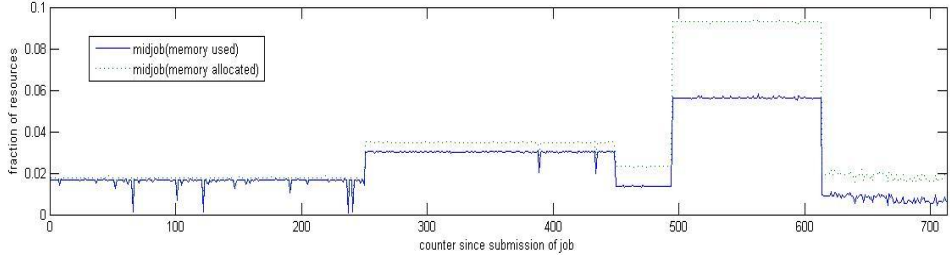

Fig. 4. Resource allocated and used in Medium Job

## C. Long Jobs

These are the jobs that are very long in duration as compared to other jobs in the trace. In fact on observing the trace it can be concluded that these are the most resource engaging jobs. Though these jobs are fewer in numbers but these are the ones that consume maximum amount of memory. Fig. 5. Corresponds to resource usage in a long job. This figure reveals that these kinds of jobs are always resource hungry and always hunting for more resources. It can also be concluded that such jobs are highly CPU intensive. Another interesting observation that was made while analysing long jobs is that there occurs symmetry amongst the tasks within each of the long jobs i.e. if one task within a long job requires a certain amount of CPU then behaviour of CPU requirements of all the other tasks within the job is also similar. Thus, we can conclude that behaviour of tasks within a long job is somewhat predictable, Fig. 6. Plots the results of a graph between memory allocated and memory used in a long job type. This figure shows that at few times more memory was allocated than required. From this we suspect that such kinds of jobs wasted a lot of memory and thus an optimal scheduling mechanism is a pre requisite. After observing the behavior of all the three jobs we can make the following observations

- **Observation 1:** Jobs in a production trace are trimodal.
- **Observation 2:** Short and medium jobs though they occur more frequently than long jobs but they utilize lesser number of resources than long jobs.
- **Observation 3:** There occurs symmetry in the tasks within a long job type.
- **Observation 4:** Behaviour of tasks within a long job type is predictable.
- **Observation 5:** Long jobs are most of the time resource hungry. by this term authors mean that jobs required most of the resources at all times and if requirement is not fulfilled, job will not be able to proceed.
- **Observation 6:** For optimal utilization of resources in production traces we propose that a blend of different job types must be used i.e. We can have a combination of different job types such as a fair number of long and short jobs or medium and short jobs.

## V. CLUSTERING OF WORKLOAD PATTERNS

Clustering is an unsupervised learning technique by which a larger group can be subdivided into several smaller groups. It can thus be used for detecting patterns within a large dataset. The advantages of this group formation technique of clustering have been used in this work for finding groupings in the trace. Identification of groups within the trace is advantageous because through this methodology we can find groupings within the jobs based on the number of resources being used by them (i.e. memory and CPU usage).This technique can further help in making important scheduling decisions such as scheduling jobs by taking jobs from different clusters based on the amount of resources being consumed by them.

### A. Clustering jobs using K-means

In order to group the available pool of jobs k-means clustering technique has been used. K-means clustering [11] is a clustering technique that partitions n observation's into k subgroups or clusters. In [10] k- means clustering has been used as a technique for management of data in a cloud based environment. Each point that belongs to a cluster is the one with minimum mean. The value of k is defined by the user. Initially the cluster mean is selected randomly and after that the algorithm iterates till point of convergence is obtained. In [2] also clustering of jobs in Google trace has been performed. In order to perform clustering, resources used by the jobs have been taken as an input vector in our approach. The mean of the amount of CPU and memory used by jobs have been taken as input vector.

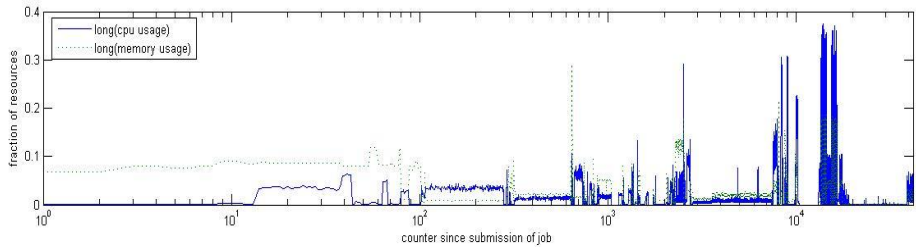

Fig. 5. Resource Usage in a Long Job

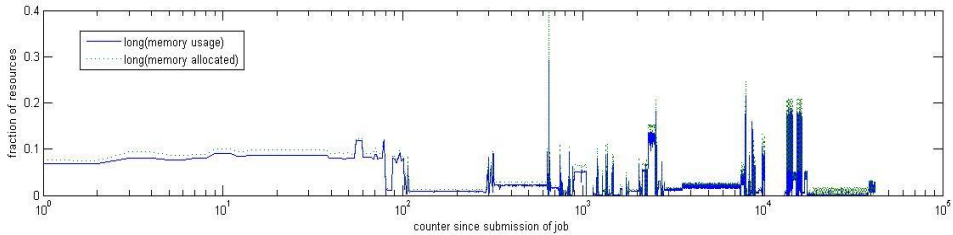

Fig. 6. Resource allocated and used in Long Job

This clustering has been performed by us using Matlab statistical toolbox. In k-means through Matlab the function returns sum point to the mean distance. The groupings amongst the different job types depends upon the number of clusters formed. Here we have used different values of k. Fig. 7. Corresponds to clustering performed on resources when clustering is performed with k=3. From this we can infer that the resources form natural groupings. This grouping can help us further validate the existence of jobs into three different types performed with k=3. Thus we can easily validate the existence of three job types short, long and medium jobs. On further iteration with k=5 as represented by fig.8, i.e. when grouping the jobs again we can see that there occurs further sub groupings amongst the different jobs. The jobs can further be classified as short jobs, approaching mid, mid, receding long and long. After clustering with k=5, it was observed that some jobs in the short job categories were found to have a pattern similar to that of short job types while others were found to have characteristics with a blend of both short and mid type, with mid type being dominant and hence were categorized as approaching mid. Mid jobs were found to retain their characteristics while in case of long jobs, a slight deviation was observed. A few of the jobs were found to retain their original characteristics while others were observed to be receding long.

Fig.9. visualizes clustering division of pool of resources, the resources can be classified into long, short and mid at k=3, they can further be classified into short, approaching mid, mid, receding long and long at k=5.

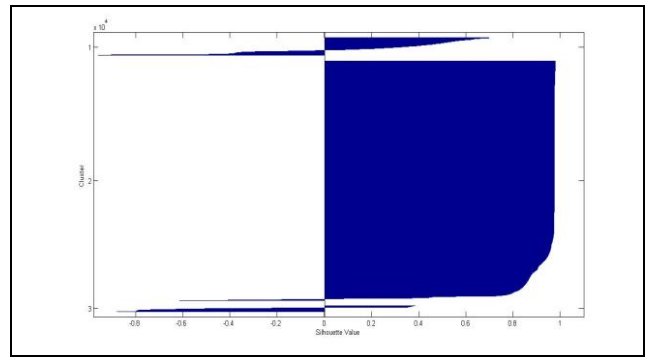

Fig. 7. Clustering with k=3

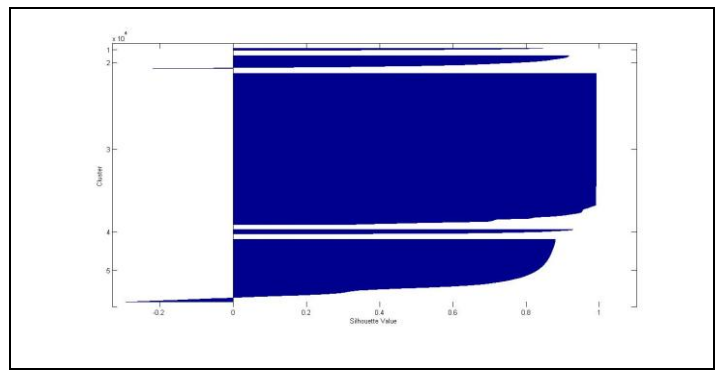

Fig. 8. Clustering with k=5

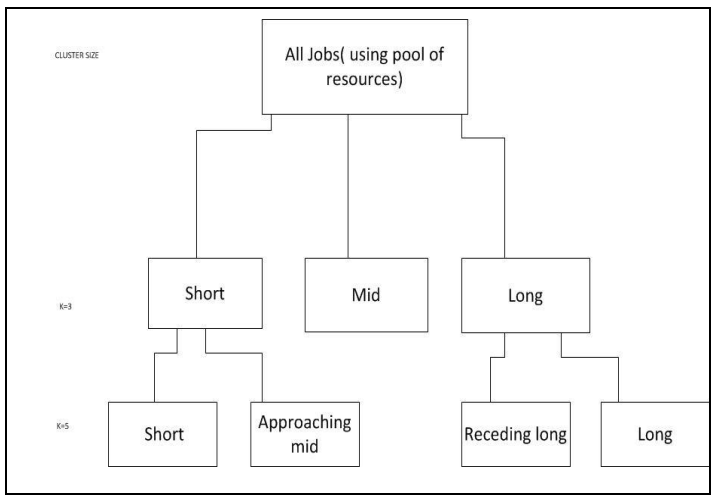

Fig. 9. K- means Clustering of Pool of Resources

## VI. CONCLUSION AND FUTURE DIRECTIONS

In order to move ahead with research in field of data management and processing of such data, we need benchmark datasets and freely available data which are publicly accessible. Even though there is a growing need for universal accessibility of such datasets but still such data is not easily available. Study of Google cluster trace is the stepping stone towards analysis of workload in a setup with multiple servers or cloud environment. This paper provides analysis of Google cluster trace by providing a statistical profile of workload behavior and clustering of jobs based on resource requirements. Though there have been earlier works for analysis of this trace, but our analysis provides several new findings such as Jobs in a production trace are trimodal and there occurs symmetry in the tasks within a long job type. In future we plan to create simulator for simulating Google and other similar production traces. We also plan to validate our findings by using Hadoop MapReduce framework on a live production server.